\documentclass[twocolumn,aps,prb,showpacs,groupedaddress]{revtex4-1}

\usepackage{graphicx}
\usepackage{amsmath}
\usepackage{amsfonts}
\usepackage{amssymb}
\usepackage{bm}

\usepackage{color}

\usepackage{xspace}
\newcommand*{\tm}{\ensuremath{T_\textrm{m}}\xspace}
\newcommand*{\units}[1]{\,\text{#1}}

\begin{document}

\title{Quantum spin coherence in halogen-modified Cr$_7$Ni molecular nanomagnets}

\author{Danielle Kaminski$^1$, Amy L. Webber$^1$, Christopher J. Wedge$^2$, Junjie Liu$^1$, Grigore A. Timco$^3$, I\~nigo J. Vitorica-Yrezabal$^3$, Eric J. L. McInnes$^3$, Richard E. P. Winpenny$^3$, Arzhang Ardavan$^1$}
\affiliation{$^{1}$Clarendon Laboratory, University of Oxford, Parks Road, Oxford OX1 3PU, U.K.} 
\affiliation{$^{2}$Department of Physics, University of Warwick, Gibbet Hill Road, Coventry CV4 7AL, U.K.} 
\affiliation{$^3$School of Chemistry and Photon Science Institute, University of Manchester, Oxford Road, Manchester M13 9PL, U.K.}
      
\date{\today}

\begin{abstract}
Among the factors determining the quantum coherence of the spin in molecular magnets is the presence and the nature of nuclear spins in the molecule. We have explored modifying the nuclear spin environment in Cr$_7$Ni-based molecular nanomagnets by replacing hydrogen atoms with deuterium or the halogen atoms, fluorine or chlorine. We find that the spin coherence, studied at low temperatures by pulsed electron spin resonance, is modified by a range of factors, including nuclear spin and {\color{black} magnetic moment}%$g$-factor
, changes in dynamics owing to nuclear mass, and molecular morphology changes.
\end{abstract}

\pacs{}

\maketitle

Since the proposal that they may be used to host elementary quantum algorithms \cite{Leuenberger2001,Tejada2001a}, research in the area of molecular magnets has increased rapidly. Such magnets offer a range of properties useful for the components of a quantum computer~\cite{AA-SJB-RSC}: flexible and tuneable electron spin structures\cite{Gatteschi2006a}; high spins permitting dense quantum memory~\cite{Leuenberger2001}; the ability to self assemble by chemical means into multiple molecular magnet `superstructures'\cite{Timco2008,Timco2009,Hill2003}; and possible formation of a dilute oriented ensemble, which allows manipulation of anisotropic spin multiplets\cite{Moro2013a}. In particular, antiferromagnetically coupled rings can form well defined spin ground states, which inherit the long coherence times of their constituent components and offer the advantage that perturbations can be applied on the molecular rather than atomic length scale~\cite{Meier2003a,Meier2003}. 

In order to be useful in a quantum computer, the phase memory time ($T_\mathrm{m}$) of a qubit must exceed significantly the duration of single-qubit manipulations. At low enough temperature, we found this to be true of a Cr$_7$Ni molecular magnet, whose paramagnetic ions can be treated as an effective electron spin-1/2 system~\cite{Ardavan2007}. By varying key structural components, the main phase decoherence mechanisms were revealed to be nuclear spin diffusion and spectral diffusion, in particular involving low mass, highly magnetic protons, both on the molecule and in the solvent. As in previous studies, methyl groups were found to be particularly effective in driving dephasing owing to the possibility of significant librational motion, and rotational tunnelling motions, even at liquid helium temperatures. By optimising the structure to reduce librational motion and replacing hydrogen with deuterium, which has an approximately six times lower magnetic moment than hydrogen, we were able to extend phase memory significantly. In a compound with all protons removed, we found $T_\mathrm{m}>15~\mu$s at 1.5~K~\cite{Wedge2012}. 

The aim of the present investigation is to determine the effects on phase memory time in Cr$_7$Ni of substituting halogens for hydrogen in the constituents of the ring. This was motivated by the chemically similar, but physically distinct, properties of these atoms (table~\ref{table:nucleardata}). Fluorine has very similar magnetic properties to hydrogen but has a mass that is nineteen times greater, enabling us to explore motional effects. In particular, trifluoromethyl (CF$_3$) groups cannot tunnel through rotational barriers whereas the three protons of conventional methyl (CH$_3$) groups are able to do so. Furthermore, the chlorine atom has a much smaller magnetic moment than hydrogen and is relatively bulky so could, for example, replace an entire methyl group, allowing construction of bulky ligands containing fewer magnetic nuclei~\cite{Eaton2000}.

\begin{table}[ht]
\caption{Nuclear spins and magetic moments of nuclei used in this study.}
\centering
\begin{tabular}{c c c}
\hline\hline
Element & Nuclear Spin & Magnetic Moment \\ 
& & Relative to $^1$H \\
\hline
$^1$H & 1/2 & 1 \\
$^2$H (D) & 1 & 0.307 \\
$^{19}$F & 1/2 & 0.941\\
$^{35}$Cl & 3/2 & 0.294 \\
$^{37}$Cl & 3/2 & 0.245 \\
\hline
\end{tabular}
\label{table:nucleardata}
\end{table}

\begin{figure}%[h] 
\centering 
\includegraphics[width=8.3cm]{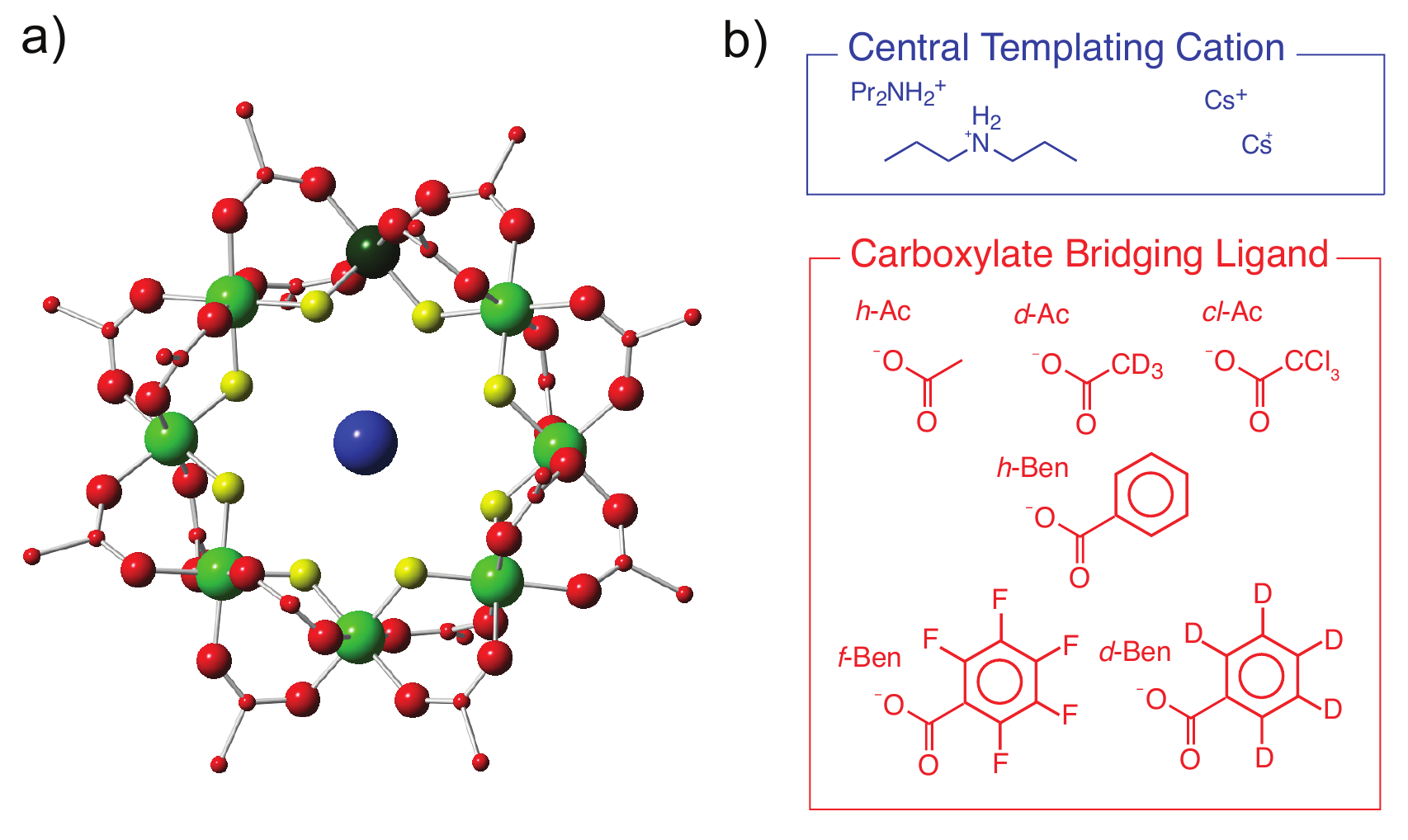} 
\caption{{\color{black} (Color online). Structures of the Cr$_7$Ni molecules.
(a)~Crystal structure of Cs[Cr$_7$NiF$_8$Ac$_{16}$]. The colored
balls represent either different atom types: Cr (light green), Ni
(dark green), F (yellow), or different interchangeable substituents:
central templating cation (blue), carboxylate bridging ligand
(red). Hydrogen atoms have been omitted for clarity. (b) Appropriately color coded chemical structures of
some possible variants with abbreviated names}%(a) The structure of a Cr$_7$Ni based single molecule magnet, with (b) the different structural components investigated
.} 
\label{fig:basicstructure} 
\end{figure}

All structures were derived from the parent compound Cr$_8$F$_8$Piv$_{16}$, which has a diamagnetic ground state. It consists of a ring of octahedrally coordinated trivalent chromium ions, each bridged to its neighbour by one fluoride and two bulky pivalate (2,2-dimethylpropanoate) carboxylate bridging ligands. Coupling between the eight Cr$^{3+}$ ($s$ = 3/2) ions is antiferromagnetic, leading to a ground state total spin, $S=0$. By substituting one Cr$^{3+}$ with a divalent Ni$^{2+}$ ($s=1$) and introducing a suitable central templating cation to balance the negative charge of the ring, we form a paramagnetic ground state of total spin, $S=1/2$. This total spin can be controlled by alternative choice of divalent metal ion~\cite{Larsen2003,Piligkos2009}, but all compounds investigated here are based on the Cr$_7$Ni ring. We explore the effect of several classes of modification: substituting the pivalate in the bridging ligand with a group in which we replace H with D, F or Cl; templating around different cations; and solvent deuteration. Figure~\ref{fig:basicstructure} shows the structures that we were motivated to study by the possibility of modifying the phase memory time of the Cr$_7$Ni spin.    

Cr$_7$Ni rings assemble readily around a wide range of cations~\cite{Wedge2012}. Here, we chose to study rings templated around either a caesium cation (Cs$^+$), which we have earlier found to support long phase coherence times, or around a large di-propyl ammonium (Pr$_2$NH$_2^+$) cation, offering an alternative synthetic target. These cations are shown in blue in Fig.~\ref{fig:basicstructure}(b). We identified acetates (Ac), and benzoates (Ben), shown in red in Fig.~\ref{fig:basicstructure}(b), as carboxylate ligands in which hydrogens may readily be exchanged for halogens. 

We found that not all combinations of cations and ligands in Fig.~\ref{fig:basicstructure} formed stable compounds. It was possible to synthesise fully Cl-substituted Ac rings around the Pr$_2$NH$_2^+$ template, but this was not stable in solution, so it was not possible to study pulsed electron spin resonance (ESR) in this structure. The analogous Cs$^+$ structure is expected to be even less stable and we did not attempt to synthesise it. Hydrogenated-, deuterated- and fluorinated-Ben ligand structures formed successfully around the Pr$_2$NH$_2^+$ cation. While it was possible to synthesise the chlorinated-Ben, it was not soluble in the solvents used for pulsed ESR spectroscopy. We were able to synthesise hydrogenated-Ben and deuterated-Ben rings around Cs$^+$, but the synthesis of fluorinated-Ben generated many by-products, and we did not obtain a sample of the pure fully-substituted ring. Details of the synthesis are given in the Supplemental Material~\cite{SM}.

\begin{figure}%[h] 
\centering 
\includegraphics[width=6.5cm]{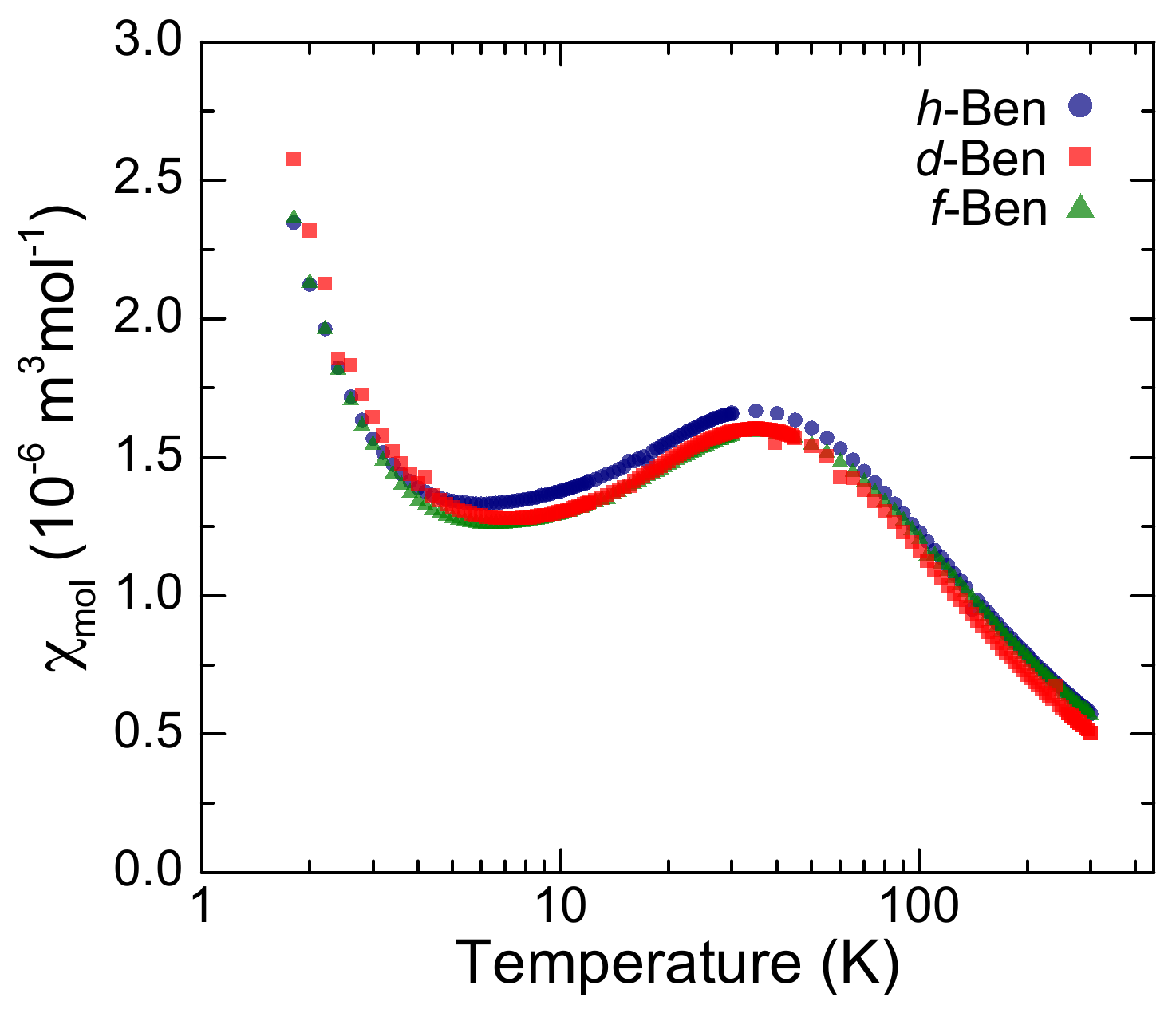} 
\caption{{\color{black} (Color online). }Magnetic susceptibility, $\chi_{\mathrm{m}}$ as a function of temperature, $T$ for compounds templated around a Pr$_2$NH$_2^+$ cation.} 
\label{fig:SQUID} 
\end{figure}  

Given that fluorine is much more electronegative than hydrogen (3.98 on the Pauling scale, compared with 2.20 for hydrogen), we considered the possibility that fluorine substitution might distort the structure sufficiently to modify the exchange couplings within the Cr$_7$Ni ring. To check this, we performed d.c.\ magnetic susceptibility measurements, using a standard Quantum Design MPMS 7~T SQUID, on powder samples of the \emph{h}-Ben, \emph{d}-Ben and \emph{f}-Ben samples templated around the Pr$_2$NH$_2^+$ cation, across the temperature range 1.8 to 300~K, with an applied magnetic field intensity of 0.1~Am$^{-1}$.
{\color{black} The susceptibility data are in agreement with previously reported values for Cr$_7$Ni~\cite{Amiri2010}, and features} %Features
in the susceptibility can be identified with magnetic excitations in the ring, permitting the extraction of the magnitude of the exchange constants~\cite{Affronte2005}. As shown in Fig.~\ref{fig:SQUID}, the magnetic susceptibilities of all three compounds are almost identical, with turning points, indicative of exchange coupling strength, occurring at the same temperatures. We conclude that there is little variation in the strength of the exchange interactions in the metal ring for the compounds that we have studied here. 

We measured phase memory times of all fully substituted, stable compounds using X-band ($\approx$~9.5~GHz) pulsed ESR spectroscopy over the temperature range 3 to 5~K, applying the standard Hahn echo sequence,  
%
%\begin{equation}\label{eqn:hahnecho}
$\pi/2 - \tau - \pi - \tau - \mathrm{echo}$,
%\end{equation}
%
%\noindent
where $\tau$ is incremented, with a two-step phase cycle~\cite{Schweiger2001}. We dissolved the compounds in dry hydrogenated or deuterated toluene (\emph{h}-tol, \emph{d}-tol respectively) and diluted such that intermolecular dipolar interactions could be neglected ($\lesssim 10^{-4}$M). We de-gassed all samples via a freeze-pump-thaw method and subsequently flame sealed the sample tubes. We flash-froze the samples before inserting them in the spectrometer to ensure that the solvent formed a glass.

\begin{figure}%[h] 
\centering 
\includegraphics[width=6.6cm]{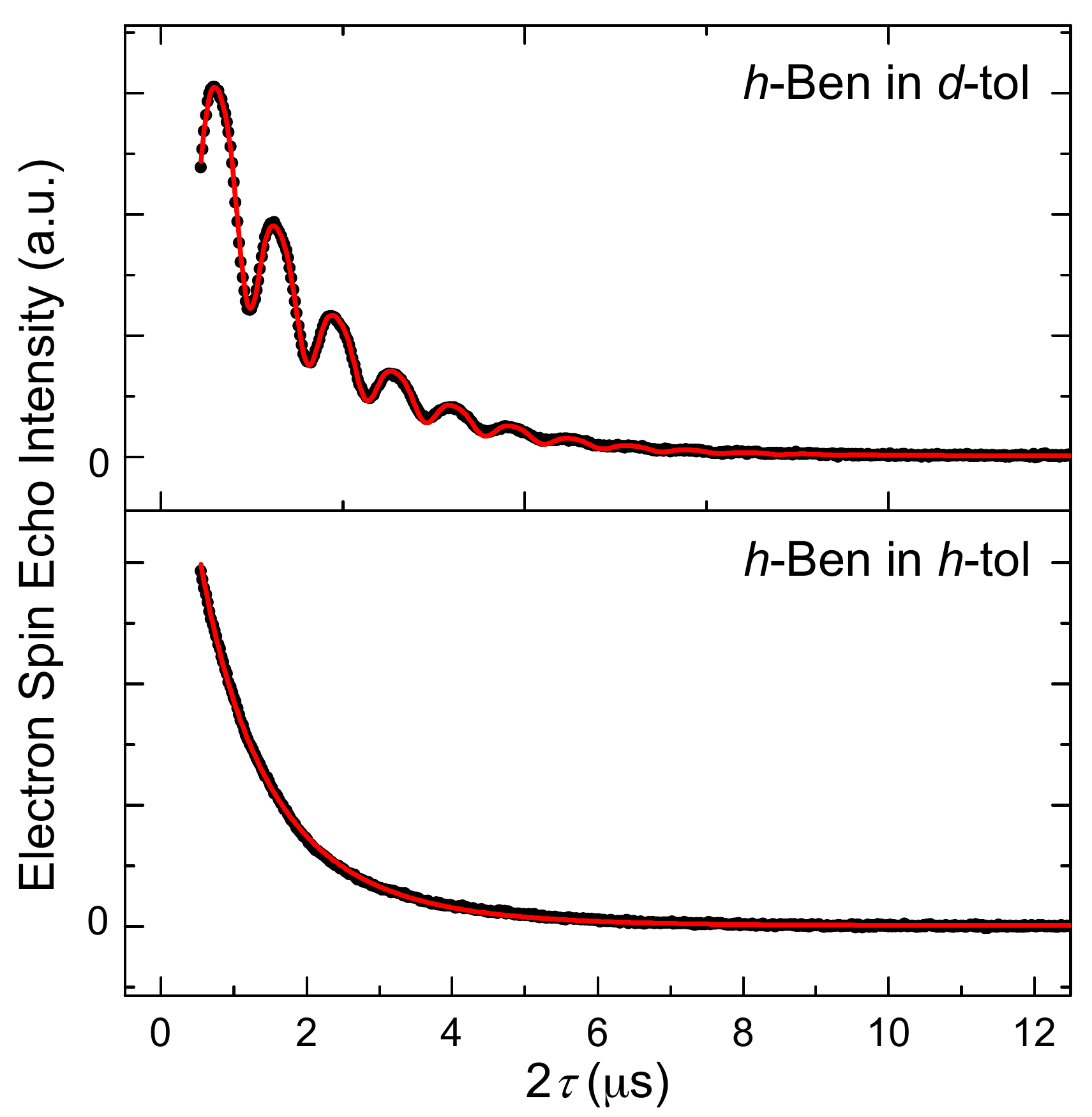} 
\caption{{\color{black} (Color online). }Echo decay curves (black points) with the relevant fits (red line), for the $h$-Ben compound templated around a Pr$_2$NH$_2^+$ cation, at temperature $T=4$~K.} 
\label{fig:examplefits} 
\end{figure}

Example spectra and fits are shown in Fig.~\ref{fig:examplefits}. In {\color{black} the top panel of} Fig.~\ref{fig:examplefits}%(a)
, there is a strong electron spin echo envelope modulation (ESEEM) at the deuterium nuclear Larmor frequency. This arises due to the excitation and subsequent interference of forbidden transitions involving a nuclear spin transition~\cite{Schweiger2001}. The magnetic moments of hydrogen and fluorine nuclei are sufficiently %low
{\color{black} large} that the bandwidth of a 128ns $\pi$-pulse is not wide enough to excite their ESEEM. Using lengthened pulses, reducing the excitation bandwidth, we were able to reduce but not to remove entirely ESEEM due to the deuterium nucleus, which has a lower magnetic moment ($\gamma_{\mathrm{H}}/\gamma_{\mathrm{D}} \approx 6.5$). Spectra not exhibiting ESEEM fitted well to a stretched exponential function (equation~\ref{eqn:stretchedexponential}) where $\tau$ is the time delay between pulses{\color{black},} %and
$Y(0)$ is the extrapolated echo intensity at $\tau=0$ {\color{black} and $c$ accounts for a small constant baseline offset of instrumental origin}. $T_{\mathrm{m}}$ and $x$ are phenomenological parameters known as the phase memory time and stretch parameter, respectively. Their values depend on both the dominant mechanism of decoherence and its rate~\cite{Eaton2000,Brown1979,Salikhov1979}. In the cases in which ESEEM was present, {\color{black} the first term of} equation~\ref{eqn:stretchedexponential} was multiplied by a modulation function accounting for both the fundamental frequency and first harmonic of the ESEEM (see Supplemental Material)~\cite{SM}. 

\begin{equation}\label{eqn:stretchedexponential}
Y(2\tau) = Y(0)\mathrm{exp}\left(-(2\tau/T_{\mathrm{m}})^x\right)+c
\end{equation}

Phase decoherence that is not refocussed in an echo experiment results from magnetic field fluctuations at the site of the excited electron spin. In the dilute, rigid limit, we expect the primary cause to be nuclear spin flip-flop processes. Although conserving net magnetisation, these cause local magnetic field fluctuations at the site of the elecron spin. The flip-flop rate of any individual nucleus is low (typically $\approx$ 10kHz), but the efficiency of the decoherence process can be enhanced by a large bath of available spins, for example, in the solvent. We refer to this process as nuclear spin diffusion and it gives rise theoretically to a stretch parameter of $2 \le x \le 3$, depending on the exact model~\cite{Eaton2000,Brown1979,Salikhov1979}. Moving away from the rigid limit, decoherence can additionally be caused by motion of magnetic nuclei with respect to the electron spin. At the temperatures investigated, we expect bond vibration to be frozen out. However rotations and librations are still possible, particularly of light atoms. We refer to decoherence resulting from these motions as spectral diffusion and this tends to give rise to a lower stretch parameter, reaching 0.5 if the motional correlation time is on the order of the pulse delay~\cite{Eaton2000}.

\begin{figure}%[h] 
\centering 
\includegraphics[width=6.6cm]{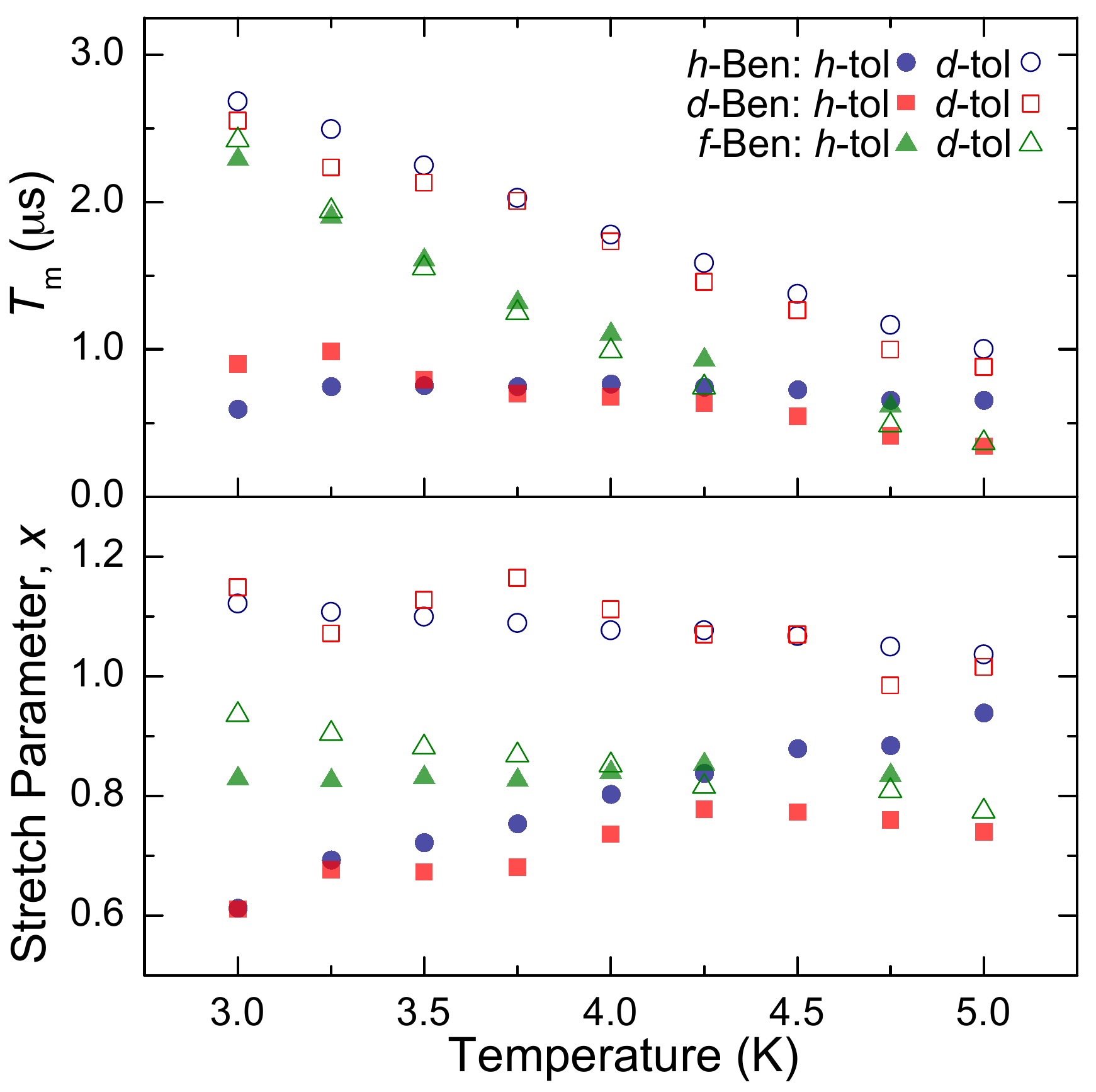} 
\caption{{\color{black} (Color online). }Phase memory time, $T_{\mathrm{m}}$ and stretch parameter, $x$ for compounds templated around a Pr$_2$NH$_2^+$ cation. Error bars from the fit are omitted as these were on the order of the marker size.} 
\label{fig:TmAmmonium} 
\end{figure}

Fig.~\ref{fig:TmAmmonium} shows the $T_{\mathrm{m}}$ and $x$ values for compounds templated around the ammonium cation. There is a marked increase in $T_{\mathrm{m}}$ at low temperatures for the fluorinated compound. As fluorine and hydrogen are magnetically similar, and we confirmed through magnetisation experiments that the intra-molecular exchange interactions are similar, we conclude that this difference arises from structural effects. As with other 2,6-substituted benzoic acids, in the pentafluorobenzoate the aromatic ring of the benzoate ligands is rotated out of the plane of the carboxyl group~\cite{Dubey2012}. This significantly alters the local environment of the central templating cation, with rotation and libration of the methyl groups significantly more hindered in the fluorinated compound. The temperature dependence of the phase memory time indicates that this motion is frozen out at lower temperatures for the fluorinated compound but continues to dominate phase decoherence through spectral diffusion for the protonated and deuterated compounds.
{\color{black} Despite variation in the efficiency of dephasing by spectral diffusion both between compounds and with temperature, in all cases the stretch parameter remains low ($x<1.2$) across the given temperature range. This indicates that spectral diffusion rather than nuclear spin diffusion is the limiting phase decoherence pathway in the regime investigated.}

We also note that the phase memory times for the $h$-Ben and $d$-Ben ligands are very similar and in \emph{h}-tol remain roughly constant with temperature. This is in contrast to previously investigated ligands such as pivalate, for which deuteration has a significant effect~\cite{Wedge2012}. The insensitivity to ligand deuteration (which is not expected to affect structure) shows that in these compounds the protons of the ligand are not involved in the limiting phase decoherence pathway. The likely reason for this is that the relatively unhindered rotation of the methyl groups of the ammonium cation is much more effective at driving phase decoherence, aided both by the possibility of quantum tunneling transitions and a shallow rotational potential well arising from the shielding of the cation from the solvent environment by the axial benzoate ligands.

\begin{figure}%[h] 
\centering 
\includegraphics[width=6.6cm]{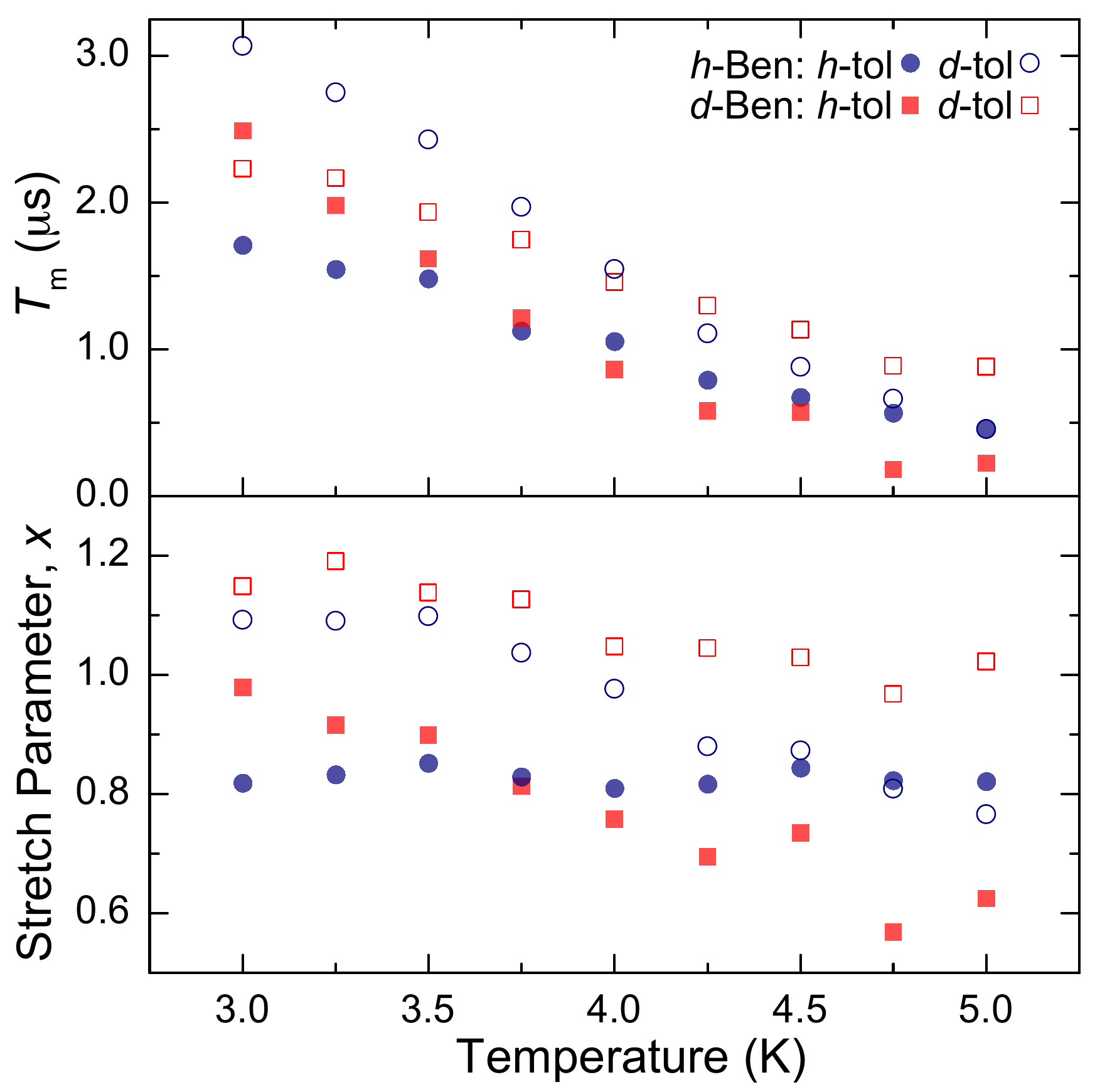} 
\caption{{\color{black} (Color online). }Phase memory time, $T_{\mathrm{m}}$ and stretch parameter, $x$ for compounds templated around a Cs$^+$ cation. Error bars from the fit are omitted as these were on the order of the marker size.} 
\label{fig:TmCs} 
\end{figure}

The $T_{\mathrm{m}}$ and $x$ values for $h$-Ben and $d$-Ben compounds templated around the Cs$^+$ cation are presented in Fig.~\ref{fig:TmCs}.
{\color{black} As in our previous study\cite{Wedge2012}, we do not observe caesium ESEEM, and we conclude that the caesium nucleus does not directly affect the phase coherence.}
Although caesium substitution of the fluorinated compound was not possible we may note that the phase memory time now increases with decreasing temperature for protonated and deuterated benzoate ligands, exceeding those in the respective ammonium templated compounds {\color{black} in \emph{h}-tol}. This supports the hypothesis that{\color{black}, when coupled to a large proton spin-bath,} the protons of the ammonium cation limit phase coherence.
{\color{black} The low stretch parameter ($x<1.2$) again indicates spectral diffusion to be the dominant decoherence mechanism in these caesium templated compounds.}

In all of the cases studied except the fluorinated compound it is found that solvent deuteration increases phase memory times. This is not unexpected, given that the solvent nuclei provide a large spin network that can be very effective in dephasing the electron spin directly, as well as acting as a spin{\color{black}-}bath for nuclear spin flip-flops with other nuclei in the system, including those of the carboxylate ligand and central templating cation.

To conclude, in order to further elucidate the decoherence mechanisms at work in Cr$_7$Ni based molecular magnets, we have explored the synthesis of a group of compounds in which hydrogen is replaced by deuterium, fluorine or chlorine. Phase memory times have been measured via pulsed ESR over the temperature range 3-5K, and we find that structural changes associated with inclusion of fluorine atoms provide the dominant effect modifying phase decoherence pathways.

\begin{acknowledgments}
This work was supported by EPSRC through the Centre for Advanced Electron Spin Resonance (CAESR), Oxford, and the National EPR Facility (Manchester).
\end{acknowledgments} 

%\bibliography{HalogenatedGreenRings}

%

%%%Supplemental material
\newpage \phantom{stuff} \newpage
\setlength{\parindent}{0cm}

{\bf\centering Supplemental Material}%
\vspace{1.5cm}

{\bf Sample preparation:} Unless stated otherwise, all reagents and solvents were purchased from commercial sources and used without further purification. Precursor Cs[Cr$_7$NiF$_8$(O$_2$C$^t$Bu)$_{16}$] was prepared similarly to its isostructural compound Cs[Cr$_7$CoF$_8$(O$_2$C$^t$Bu)$_{16}$] %\cite{faust}
[1]. Analytical data was obtained by the Microanalytical Service at the University of Manchester. \\

{\bf Precursor [$^n$Pr$_2$NH$_2$][Cr$_7$NiF$_8$(O$_2$CC$_2$H$_5$)$_{16}$] (1):} Propionic acid (75\units{mL}, 1005.33\units{mmol}) $n$-dipropylamine (4.0\units{g}, 39.53\units{mmol}), chromium(III) fluoride tetrahydrate (25.0\units{g}, 138.08\units{mmol}) and nickel(II) carbonate hydroxide tetrahydrate (4.0\units{g}, 6.81\units{mmol}) were combined in a Teflon flask and stirred for 50\units{hours} at 150$^\circ$C. After 20\units{hours} an additional 25\units{mL} of propionic acid was added. The solution was cooled, after which diethyl ether (100\units{mL}) was added, the solution stirred for 30\units{minutes} then placed on a silica column with diethyl ether as the solvent. The column was then run with diethyl ether to elute a dense green band, which was collected in 300 mL of diethyl ether. To the eluted solution was added heptane (100\units{mL}) and the solvent slowly removed under reduced pressure to completely remove diethyl ether and partially remove heptane. Precipitate formed and to the slurry was added petroleum ether 40-60$^\circ$C (300\units{mL}) and stirred overnight. The resultant precipitate was collected by filtration, washed with petroleum ether 40-60$^\circ$C (5 $\times$ 50\units{mL}) and dried. Yield: 27.6\units{g}, 14.95\units{mmol} (75.8\% based on CrF$_3$.$4$H$_2$O). Elemental analysis (\%) calc. for C$_{54}$H$_{96}$Cr$_{7}$F$_{8}$NNiO$_{32}$: C 35.13 H 5.24 N 0.76 Cr 19.72 Ni 3.18; found: C 35.30 H 5.24 N 0.72 Cr 19.72 Ni 3.10. \\

{\bf Compound [$^n$Pr$_2$NH$_2$][Cr$_7$NiF$_8$(O$_2$CC$_6$H$_5$)$_{16}$] (2):} Chromium(III) fluoride tetrahydrate (1.0\units{g}, 5.52\units{mmol}), nickel(II) carbonate hydroxide tetrahydrate (0.25\units{g}, 0.43\units{mmol}), benzoic acid (5.0\units{g}, 40.94\units{mmol}), dipropylamine (0.35\units{g}, 3.46\units{mmol}) and 1,2-dichlorobenzene (5\units{ml}) were stirred together in a Teflon flask at 160$^\circ$C for 22\units{h}. The flask was then allowed to cool to room temperature and acetone (20\units{ml}) was added and stirred for 30\units{min}, then the solid  was filtered, and washed with acetone. After this the solid was stirred with dichloromethane (500\units{ml}) for \emph{ca}.~30\units{min} and the solution obtained was filtered. The solvent from the filtrate was removed under reduced pressure leaving a green residue that was washed with acetone and dried \emph{in vacuo.} Yield: 1.44\units{g} (70\%, based on Cr). Elemental analysis calculated (\%) for C$_{118}$H$_{96}$Cr$_{7}$F$_{8}$N$_{1}$Ni$_{1}$O$_{32}$: Cr 13.92, Ni 2.24, C 54.20, H 3.70, N 0.54; found: Cr 13.76, Ni 2.40, C 54.19, H 3.26, N 0.55. X-Ray quality crystals were obtained from the recrystallization of \textbf{2} from a mixture of DCM/Toluene/MeCN. \\

{\bf Compound [$^n$Pr$_2$NH$_2$][Cr$_7$NiF$_8$(O$_2$CC$_6$D$_5$)$_{16}$] (3):} Compound [$^n$Pr$_2$NH$_2$][Cr$_7$NiF$_8$(O$_2$CC$_2$H$_5$)$_{16}$] (0.5\units{g}, 0.27\units{mmol}), benzoic acid-2,3,4,5,6-d5 (99 atom \% D) (2.0\units{g}, 15.73\units{mmol}) and 1,2-dichlorobenzene anhydrous (4.0\units{ml}) were stirred together in a Teflon flask at 160$^\circ$C for 26\units{h} in a flow of N$_2$. Then the flask was allowed to cool to room temperature and acetone (20\units{mL}) was added and stirred for 30\units{min}, then the solid was filtered, and washed with acetone. After this the solid was stirred with dichloromethane (100\units{ml}) for \emph{ca}.~30\units{min} and the solution obtained was filtered. The solvent from the filtrate was removed under reduced pressure leaving a green residue that was washed with acetone and dried \emph{in vacuo.}  Yield: 0.6\units{g} (82\%). Elemental analysis calculated (\%) for C$_{118}$H$_{16}$N$_1$Cr$_7$Ni$_1$F$_8$O$_{32}$D$_{80}$: Cr 13.50, Ni 2.18, C 52.58, N 0.52; found: Cr 13.33, Ni 2.14, C 52.59, N 0.55. X-Ray quality crystals were obtained from the recrystallization of \textbf{3} from a  mixture of DCM/Toluene/MeCN. \\

{\bf Compound Cs[Cr$_7$NiF$_8$(O$_2$CC$_6$H$_5$)$_{16}$] (4):} Compound Cs[Cr$_7$NiF$_8$(O$_2$C$^t$Bu)$_{16}$] (1.0\units{g}, 0.43\units{mmol}), benzoic acid (5.0\units{g}, 40.98\units{mmol}) and diethylene glycol dimethyl ether anhydrous (20\units{ml}) were stirred together in a Teflon flask at 165$^\circ$C for 30\units{h} in a flow of N$_2$. Then the flask was allowed to cool to room temperature and acetone (25\units{ml}) was added and stirred for 15\units{min} then the solid was filtered, and washed with acetone. After this the solid was stirred with dichloromethane (100\units{ml}) for \emph{ca}.~30\units{min} and the solution obtained was filtered. The solvent from the filtrate was removed under reduced pressure leaving a green residue that was washed with diethyl ether and dried \emph{in vacuo.}  Yield: 0.77\units{g} (68\%). Elemental analysis calculated (\%) for C$_{112}$H$_{80}$Cr$_7$Cs$_1$F$_8$Ni$_1$O$_{32}$: Cr 13.76, Ni 2.22, C 50.85, H 3.05; found: Cr 13.25, Ni 2.10, C 51.3. X-Ray quality crystals were obtained from the recrystallization of \textbf{4} from a mixture of DCM/Toluene. \\

{\bf Compound Cs[Cr$_7$NiF$_8$(O$_2$CC$_6$D$_5$)$_{16}$] (5):} Compound Cs[Cr$_7$NiF$_8$(O$_2$C$^t$Bu)$_{16}$] (0.6\units{g}, 0.26\units{mmol}), benzoic acid-2,3,4,5,6-d5 (99 atom \% D) (2.0\units{g}, 15.73\units{mmol}) and diethylene glycol dimethyl ether anhydrous (10\units{ml}) were stirred together in a Teflon flask at 165$^\circ$C for 30\units{h} in a flow of N$_2$. Then the flask was allowed to cool to room temperature and acetone (20\units{ml}) was added and stirred for 15\units{min} then the solid was filtered, and washed with acetone. After this the solid was stirred with dichloromethane (100\units{ml}) for \emph{ca}.~30\units{min} and the solution obtained was filtered. The solvent from the filtrate was removed under reduced pressure leaving a green residue that was washed with diethyl ether and dried \emph{in vacuo.} Yield: 0.5\units{g} (71\%). Elemental analysis calculated (\%) for Cs$_1$Cr$_7$Ni$_1$F$_8$O$_{32}$C$_{112}$D$_{80}$: Cr 13.35, Ni 2.15, C 49.35; found: Cr 13.18, Ni 1.99, C 49.03. X-Ray quality crystals were obtained from the recrystallization of \textbf{5} from a mixture of DCM/Toluene. \\

{\bf Compound [$^n$Pr$_2$NH$_2$][Cr$_7$NiF$_8$(O$_2$CC$_6$F$_5$)$_{16}$] (6):} Compound [$^n$Pr$_2$NH$_2$][Cr$_7$NiF$_8$(O$_2$CC$_2$H$_5$)$_{16}$] (3.6\units{g}, 1.95\units{mmol}), 2,3,4,5,6-pentafluorobenzoic acid (10.0\units{g}, 47.15\units{mmol}), 1,2-dichlorobenzene anhydrous (5\units{ml}) and toluene anhydrous were refluxed under N$_2$ for 30\units{h}, then the solvents were removed by distillation by increasing the temperature of oil bath up to 170$^\circ$C. Then the flask was cooled to 140$^\circ$C and kept for 5\units{h} under a flow of N$_2$. Pentafluorobenzoic acid was partially sublimed during this time. After this the solid was stirred under reflux with 1,4-dioxane anhydrous (160\units{ml}) for \emph{ca}.~30\units{min} and the solution obtained was filtered hot. The solution was kept at room temperature for two weeks under N$_2$. During this time a microcrystalline (including X-ray quality crystals) green product was precipitated. Product was filtered, washed with hexane anhydrous and dried \emph{in vacuo.} Yield 2.35 g (30 \%). Elemental analysis calculated (\%) for C$_{118}$H$_{16}$Cr$_7$F$_{88}$N$_1$Ni$_1$O$_{32}$: Cr 8.98, Ni 1.45, C 34.96, H 0.40, N 0.35; found: Cr 9.07, Ni 1.35, C 35.19, H 0.2, N 0.36. \\

{\bf Compound [$^n$Pr$_2$NH$_2$][Cr$_7$NiF$_8$(O$_2$CCCl$_3$)$_{16}$] (7):} Compound [$^n$Pr$_2$NH$_2$][Cr$_7$NiF$_8$(O$_2$CC$_2$H$_5$)$_{16}$] (2.0\units{g}, 1.08\units{mmol}), trichloroacetic acid (5.0\units{g}, 30.60\units{mmol}), and o-xylene anhydrous (50\units{ml}) were refluxed for 20\units{h} under N$_2$, then the solvent was removed and fresh o-xylene (30\units{ml}) added and solution refluxed while stirring for another 5\units{h}, then the solvent was removed under reduced pressure and residue kept under a flow of N$_2$ for 3\units{h} at 120$^\circ$C. After this the flask was cooled to room temperature and solid was extracted under reflux while stirring with hexane anhydrous (3 $\times$ 250\units{ml}). The solution obtained was filtered hot and concentrated by distillation to \emph{ca.}~200\units{ml}, and was kept at room temperature for two days under N$_2$. During this time a microcrystalline (including X-ray quality crystals) green product was precipitated. Product was collected by filtration, washed with hexane and dried \emph{in vacuo.} Yield: 0.65\units{g} (18\%). Elemental analysis calculated (\%) for C$_{38}$H$_{16}$Cl$_{48}$Cr$_7$F$_8$N$_1$Ni$_1$O$_{32}$: Cr 11.11, Ni 1.79, C 13.94, H 0.49, N 0.43; found: Cr 11.53, Ni 1.89, C 14.08, H 0.40, N 0.36. \\

{\bf Crystallographic data:} The X-ray data for single crystals of compounds \textbf{2} and \textbf{3} were collected using Cu-K$_{\alpha}$ radiation on an X8 Bruker Prospector, for compound \textbf{5} using Mo-K$_{\alpha}$ radiation on an Agilent Supernova and for compounds \textbf{6} and \textbf{7} using synchrotron radiation at Diamond Light Source
%\cite{Diamond}
[2]. A partial data set was collected on compound \textbf{4}, which was found to be isostructural with compound \textbf{5}; the crystals diffracted poorly so a full structural determination was not possible. 

The crystals of compound \textbf{6} were twinned, and it was modeled as a four-component twin. The crystals of compound \textbf{7} suffered from significant beam damage and therefore the data are only 93\% complete. 

All structures suffer from various degrees of crystallographic disorder, which have been modeled using standard techniques and programs~[3,4];
%\cite{sheldrick,dolomanov}; 
the models adopted are explained in detail in the deposited crystallographic data files. Table~\ref{crystallographic_info} summarises crystallographic information for compounds \textbf{2}--\textbf{3}, and \textbf{5}--\textbf{7}. Full crystallographic data can be found in CIF format on the Cambridge Structural Database: CCDC numbers 1022266 - 1022270. \\

\begin{turnpage}
\begin{table}[htb]
\caption{Table of crystallographic information for compounds \textbf{2} - \textbf{4}, \textbf{6} and \textbf{7}.}
\label{crystallographic_info}
\centering
\begin{tabular}{c c c c c c}
\hline\hline
 & \textbf{2} & \textbf{3} & \textbf{5}  & \textbf{6} & \textbf{7} \\
\hline
Chemical Formula & C$_{118}$H$_{96}$Cr$_7$F$_8$NNiO$_{34}$ & C$_{120}$H$_{16}$D$_{83}$Cr$_7$F$_8$N$_2$NiO$_{32}$ & C$_{112}$D$_{80}$Cr$_7$CsF$_8$NiO$_{34}$ & C$_{150}$H$_{72}$Cr$_7$F$_{88}$NNiO$_{48}$ & C$_{44}$H$_{28}$Cl$_{48}$Cr$_7$F$_8$NNiO$_{32}$ \\
Relative Molar Mass&2646.66&2739.23&2757.87&4750.79&3358.98 \\
Temperature/K&100(2)&100(2)&150.0(2)&100(2)&100(2) \\
Crystal System &Tetragonal&Tetragonal&Tetragonal&Tetragonal&Monoclinic \\
Space Group &$P4/nnc$&$P4/nnc$&$P4/nnc$&$P42_12$&$P2_1/c$ \\
$a/$\AA &16.6109(3)&16.5043(4)&16.4968(6)&23.648(3)&16.426(1) \\
$b/$\AA &16.6109(3)&16.5043(4)&16.4968(6)&23.648(3)& 29.293(1) \\
$c/$\AA &23.2263(8)&23.225(1)&23.128(2)&17.780(3)& 25.165(1) \\
$\beta/^{\circ}$ &90&90&90&90&106.080(6) \\
$V/$\AA$^3$ &6408.6(3)&6326.1(4)&6294.1(6)&9943(3)&11661(1) \\
$Z$ &2&2&2&2&4 \\
$\lambda/$\AA &1.5418&1.5418&0.71073&0.6889&0.6889 \\
$\rho$ calc'd/g cm$^{-3}$ &1.372&1.446&1.455&1.587&1.913 \\
$\mu$ (Mo K$_{\alpha}$)/mm$^{-1}$ &5.587&5.652&1.095&0.614&1.808 \\
$R_1(I>2\sigma(I))\,^a$ &0.0820&0.0654&0.0902&0.0993&0.1703 \\
$wR_2\,^b$ &0.2420&0.1870&0.2050&0.2393&0.4911 \\
\hline
\multicolumn{6}{l}{$^aR_1 = ||F_\textrm{o}| - |F_{\textrm{c}}||/|F_{\textrm{o}}|$} \\
\multicolumn{6}{l}{$^bwR_2 = \left[ w(|F_\textrm{o}| - |F_{\textrm{c}}|)^2/w|F_{\textrm{o}}|^2\right]^{1/2}$}
\end{tabular}
\end{table}
\end{turnpage}

{\bf Data fitting:} Phase memory times were determined by least squares fitting of the echo decay with a stretched exponential using a solver based on the Trust-Region algorithm. The fitting function used was $Y(2\tau)=Y(0)\exp(-(2\tau/\tm)^x)+c$ or $Y(2\tau)=Y(0)\exp(-(2\tau/\tm)^x)(1+k_1\sin(2\omega\tau+\phi_1)+k_2\sin(4\omega\tau+\phi_2))+c$ for strongly modulated data, with $Y(0)$, $\tm$, $x$, $k_1$, $k_2$, $\phi_1$, $\phi_2$, $\omega$ and $c$ being freely varying fitting parameters. This approach is appropriate as in the weak coupling regime only the nuclear frequency ($\omega_I$) and its harmonic ($2\omega_I$) are observed in the ESEEM spectrum. Note that the inclusion of the harmonic contribution only adds two extra variable parameters, the modulation depth $k_2$ and phase $\phi_2$, as the frequency is constrained by the parameter used for the fundamental. To reduce the number of variable parameters the modulations have not been permitted to decay independently from the overall decay of the echo signal, although in reality the decay of modulations may be faster [5].
% \cite{jeschke_book}.

%%%%%%%%%%%%%%%% Bibliography %%%%%%%%%%%%%%%%%%%%%%%%%%%%%%%%%%%

%
% APS
% First Author et al. allowed "Only if length constrained and four or more authors"
%

%\bibliographystyle{apsrev4-1}
%\begin{thebibliography}{0}

%\bibitem{wedge:prl2012}
%C.J.~Wedge \textit{et al.},
%%Chemical Engineering of Molecular Qubits
%Phys. Rev. Lett., \textbf{108}, 107204 (2012).
%%http://dx.doi.org/10.1103/PhysRevLett.108.107204

%\bibitem{faust}
\vspace{0.3cm} [1] 
T. B.~Faust \textit{et al.},
% Caesium ion sequestration by a fluoro-metallocrown [16]-MC-8
Chem. Commun., \textbf{46}, 6258 (2010).
%http://dx.doi.org/10.1039/C0CC01188F

%\bibitem{Diamond}
\vspace{0.3cm} [2] 
H. Nowell \textit{et al.},
J. Synchrotron Rad., \textbf{19}, 435 (2012).

%\bibitem{sheldrick}
\vspace{0.3cm} [3] 
G. M. Sheldrick,
Acta Cryst. Sec. A \textbf{64}, 112 (2008).

%\bibitem{dolomanov}
\vspace{0.3cm} [4] 
O. V. Dolomanov \textit{et al.},
J. Appl. Cryst. \textbf{42}, 339 (2009).

%\bibitem{jeschke_book}
\vspace{0.3cm} [5] 
A.~Schweiger and G.~Jeschke,
\newblock {\em Principles of Pulsed Electron Paramagnetic Resonance}
\newblock (Oxford Univ. Press, Oxford, 2001).

%\end{thebibliography}

%%%%%%%%%%%%%%%%%%%%%%%%%%%%%%%%%%%%%%%%%%%%%%%%%%%%%%%%%%%%%

\end{document}